\documentclass{article}
\pdfoutput=1
\usepackage{graphicx}        
\usepackage{spconf,amsmath,graphicx}
\usepackage{subfig}
\usepackage{multirow}
\usepackage{rotating}
\usepackage{amsfonts, cite}
\usepackage{array}
\usepackage{float}
\usepackage{comment}
\usepackage{booktabs}
\newcolumntype{C}[1]{>{\centering\arraybackslash}p{#1}}

\title{Towards seamless authentication for Zoom-based online teaching and meeting}
\name{Manoranjan Mohanty$^{\odot}$ and Waheeb Yaqub $^{\star}$ 
}
\address{
$^{\odot}$Center for Forensic Science, University of Technology Sydney, Australia
\\
$^{\star}$School of Computer Science, The University of Sydney, Australia\\
}

\begin{document}
\ninept
\maketitle

\begin{abstract}
The lockdowns and travel restrictions in current coronavirus pandemic situation has replaced face-to-face teaching and meeting with online teaching and meeting. Recently, the video conferencing tool Zoom has become extremely popular for its simple-to-use feature and low network bandwith requirement. However, Zoom has serious security and privacy issues. Due to weak authentication mechanisms, unauthorized persons are invading Zoom sessions and creating disturbances (known as Zoom bombing). In this paper, we propose a preliminary work towards a seamless authentication mechanism for Zoom-based teaching and meeting. Our method is based on PRNU (Photo Response Non Uniformity)-based camera authentication, which can authenticate the camera of a device used in a Zoom meeting without requiring any assistance from the  participants (e.g., needing the participant to provide biometric). Results from a small-scale experiment validates the proposed method. 

\end{abstract} 


\section{Introduction}
\label{sec:Introduction}
The current coronavirus (covid-19) pandemic situation has brought a lot of changes in how teaching and meeting are happening around the world. Due to lockdowns and travel restrictions, regular face-to-face
teaching (classroom teaching) and face-to-face meeting are being replaced with video conference-based online teaching and meeting. Recently, the video conferencing tool Zoom\footnote{https://zoom.us/} has become extremely popular for its simple-to-use feature and low network bandwidth requirement, resulting so-called \textit{Zoom booming}~\cite{zoom_boom}. 

However, as has been reported by many media houses and acknowledged by Zoom, Zoom has serious security and privacy issues~\cite{zoom_security}. For lowering the network bandwidth and network latency requirements, Zoom does not use end-to-end encryption (as encryption introduces extra overheads). Although a password-based authentication mechanism has been provided, the use of password is also optional for making Zoom more user-friendly. The password setting is also not provided in the default Zoom setup. As a result, many Zoom sessions are password-less. Some Zoom sessions are attended by an unauthorized persons, leading to \textit{Zoom bombing} (teachers being racially abused and students being shown pornographic videos), and \textit{Zoom eavesdropping} (confidential conversation being secretly heard) issues~\cite{zoom_porn, zoom_porn1}. 

A combination of Zoom features can be used for addressing the security and privacy issues in some extent~\cite{zoom_sol1}. Besides using the password, the \textit{wait room} feature can be used by the teacher or the meeting host for accepting or rejecting joining requests from  students or meeting participants~\cite{zoom_sol2}. Although this feature can be useful in controlling a small classroom or meeting, for a bigger classroom or meeting (where hundreds of participant can join), this feature can also fail. It could be near impossible for the teacher or the meeting host for remembering hundreds of names and online identities of students and meeting participants. This solution also can create a lot of hassle as some of the joining requests can be attended in-the-middle of the Zoom session. Besides the \textit{wait room} feature, other guidelines, such as not to share the meeting links in public domain, sharing the meeting links just before the start of the meeting and not using recurring meeting options, and controlling the screen share option etc. can be used. None of them, however, can provide full-proof defense against a savvy attacker (e.g., a hacker). 

By taking Zoom as an example, in this paper, we present a preliminary work towards a  scalable, automatic, and hassle-free authentication scheme for video-conferencing-based online meeting and teaching. The proposed scheme combines PRNU~\cite{Lukas:PRNU:2006} (Photo Response Non-Uniformity)-based authentication with password-based authentication. The PRNU-based authentication can seamlessly authenticate a meeting participant (or a student) by authenticating the camera of the device she uses in the meeting (or online classroom). In such cases, the meeting participant enjoys the same usability that she enjoys with default zoom setup (no password to remember, no keys to press, no bio-metric required). Whenever the PRNU-based scheme cannot authenticate, the participant is asked to enter the password. The proposed scheme was experimentally validated by validating the effectiveness of the PRNU-based method for the webcams of desktops, and selfie-camera of mobile phones and tablets (as these cameras are typically used in Zoom sessions). The experiment with a set of $10$ cameras shows promising result. 

The rest of the paper has been organized as follows. Section~\ref{sec:RelatedeWork1} presents related work and provides an overview of how Zoom works. In Section~\ref{sec:OurApproach}, we discuss the proposed method. Section~\ref{sec:experiments} presents experimental result. Finally, Section~\ref{sec:ConclusionAndFutureWork} concludes and also discusses how we intend to extend this paper.

In the rest of the paper, we will use the term Zoom meeting in the place of Zoom-based meeting or Zoom-based teaching.

\section{Background and Related Work}
\label{sec:RelatedeWork1}
\subsection{Video Conferencing Using Zoom}
Zoom is arguably one of the most popular video conferencing tool now. One of the main reasons for this popularity is Zoom's easy-to-use feature. For 
joining an online meeting room, the participant does not need to go with the hassle of registering into Zoom. Only a meeting ID (which is a multi-digit number) is enough. By default, Zoom does not authenticate a participant. This leads to a number of security and privacy issues, such as authentication issue. 

\begin{figure}
    \centering
    \includegraphics[width=2.5in]{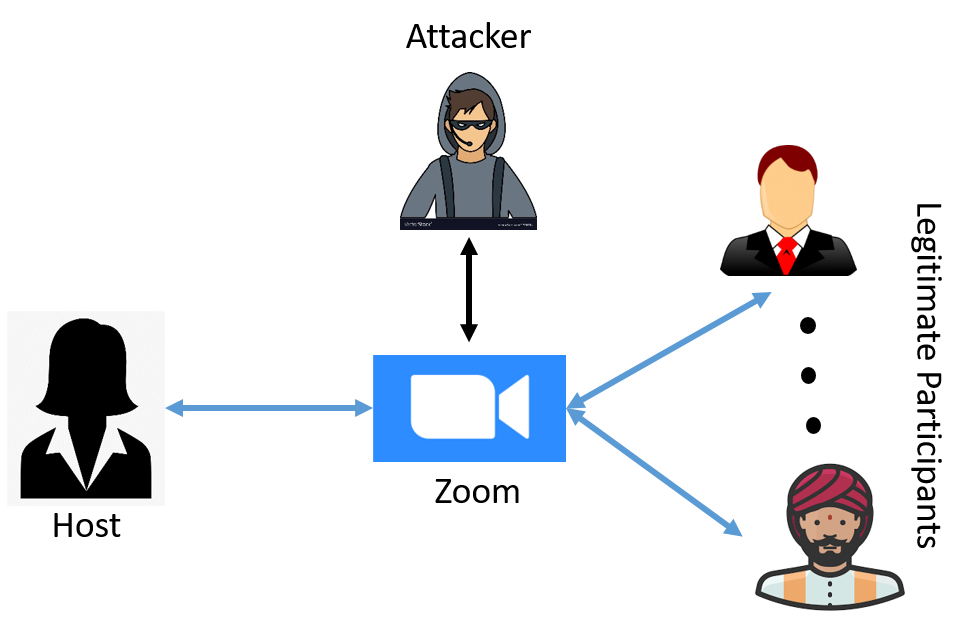}
    \caption{Threat Model.}
    \label{fig:threatmodel}
\end{figure}
Figure~\ref{fig:threatmodel} shows how authentication threat can happen for a Zoom meeting with default meeting setup. It is assumed that the Attacker can know the meeting ID. This is a safe assumption as people are sharing meeting IDs in public forum (e.g., social media). It is also assumed that the host and legitimate participants will not often detect that the Attacker has joined. This is also a safe assumption as the attacker can join in-between the meeting when the host and participants are busy. After joining the meeting, the attacker can launch Zoom bombing and Zoom eavesdropping.

After a series of security issues, Zoom has encouraged to use its wait room feature. Although this feature can address authentication issue for a small meeting,  authentication is still a concern for a large meeting.  For a large meeting, it would be difficult for the host to remember the identities of all participants (e.g., hundreds of name). In such case, it is safe to assume that either the host will not use the wait room feature or randomly approve connections requests. This will make the threat model presented in Figure~\ref{fig:threatmodel} valid.

\begin{figure}
    \centering
    \includegraphics[width=3.5in]{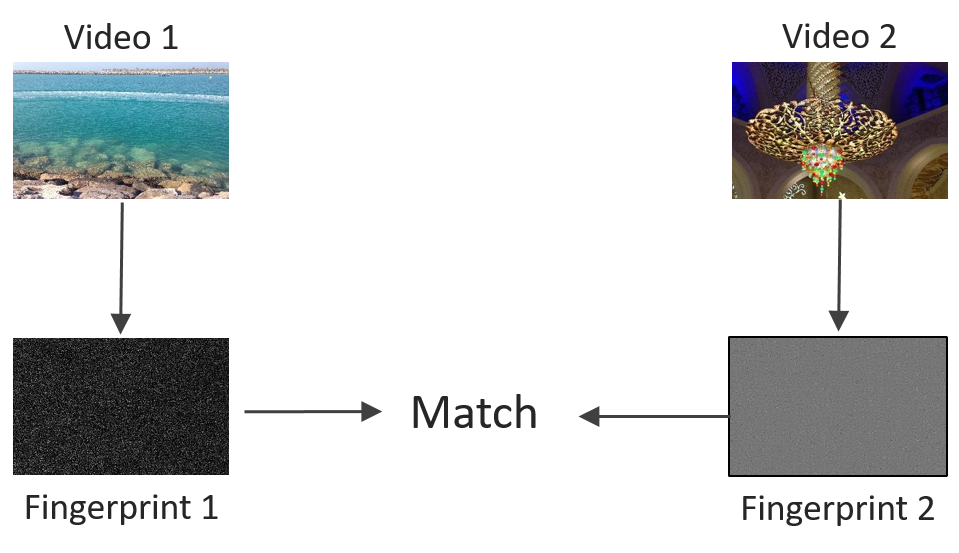}
    \caption{Camera attribution using two videos.}
    \label{fig:my_label1}
\end{figure}

\subsection{PRNU-Based Camera Fingerprinting}
The PRNU (Photo Response Non-Uniformity)-based source camera attribution is an effective method for
identifying if two different videos (or images) belong to the same camera~\cite{Lukas:PRNU:2006, Samet:Video:2016, Chen:2007:Vid, jessica:camcorder}. Figure~\ref{fig:my_label1} shows how this method works. 
PRNU-based camera attribution is based on the fact that the output of the camera sensor, $I$, can be modeled as 
\begin{equation}
 I = I^{(0)} + I^{(0)}K + \psi
\label{Eq:im_noise}
\end{equation}
where $I^{(0)}$ is the noise-free video frame (or an image), $K$ is the PRNU noise, and $\psi$ is the combination of additional noise, such as readout noise, dark current, and quantization noise. The multiplicative PRNU noise pattern, $K$, is unique for each camera and can be used as a camera fingerprint which enables the attribution of a video to its source camera.
Using a denoising filter $F$ (such as a Wavelet filter) on a set of video frames of a camera (where it is known that the video belongs to the camera, physical access to the camera is not required),
we can estimate a known camera fingerprint by first getting the
noise residual, $W_k$, (i.e., the estimated PRNU) of the $k^{th}$ frame as $W_k = I_k - \hat{I}^{(0)}_k, \hat{I}^{(0)}_k = F(I_k)$, and then averaging the noise residuals of all the frames.
For determining if a specific camera has taken
a given query video, we similarly obtain a query fingerprint and the match (correlate) this fingerprint with the known fingerprint. 
The matching is typically done using Peak-to-Correlation Energy (PCE) with a matching threshold of $60$. If the matching score is above the threshold, it is concluded that both videos belong to the same camera. So far, this PRNU method has been used for camera verification, camera identification, image/video clustering, etc. 

The PRNU-based method has also been used for authentication of users via the camera they use~\cite{Val:2017:PRNUAuth, Ba:2018:ABC}. The existing schemes, however, have been designed for images. In contrast, our method considers video. Also, unlike previous schemes, our method
combines the PRNU-based method with password.

\section{Proposed Method}
\label{sec:OurApproach}
\begin{figure}
    \centering
    \includegraphics[width=3.5in]{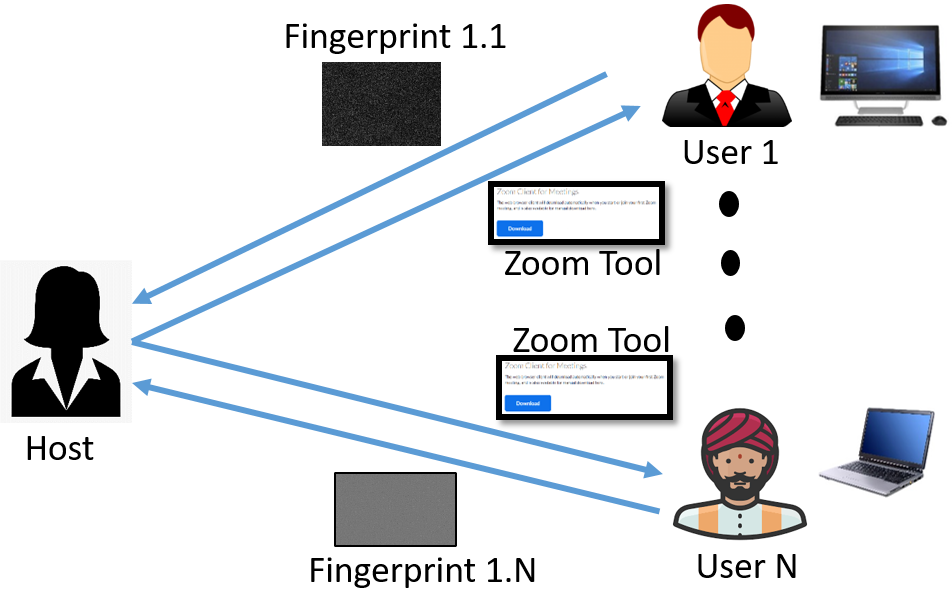}
    \caption{Camera fingerprint registration. 'User 1' to 'User N' are N legitimate participants.}
    \label{fig:Regi}
\end{figure}

\begin{figure}
    \centering
    \includegraphics[width=3.5in]{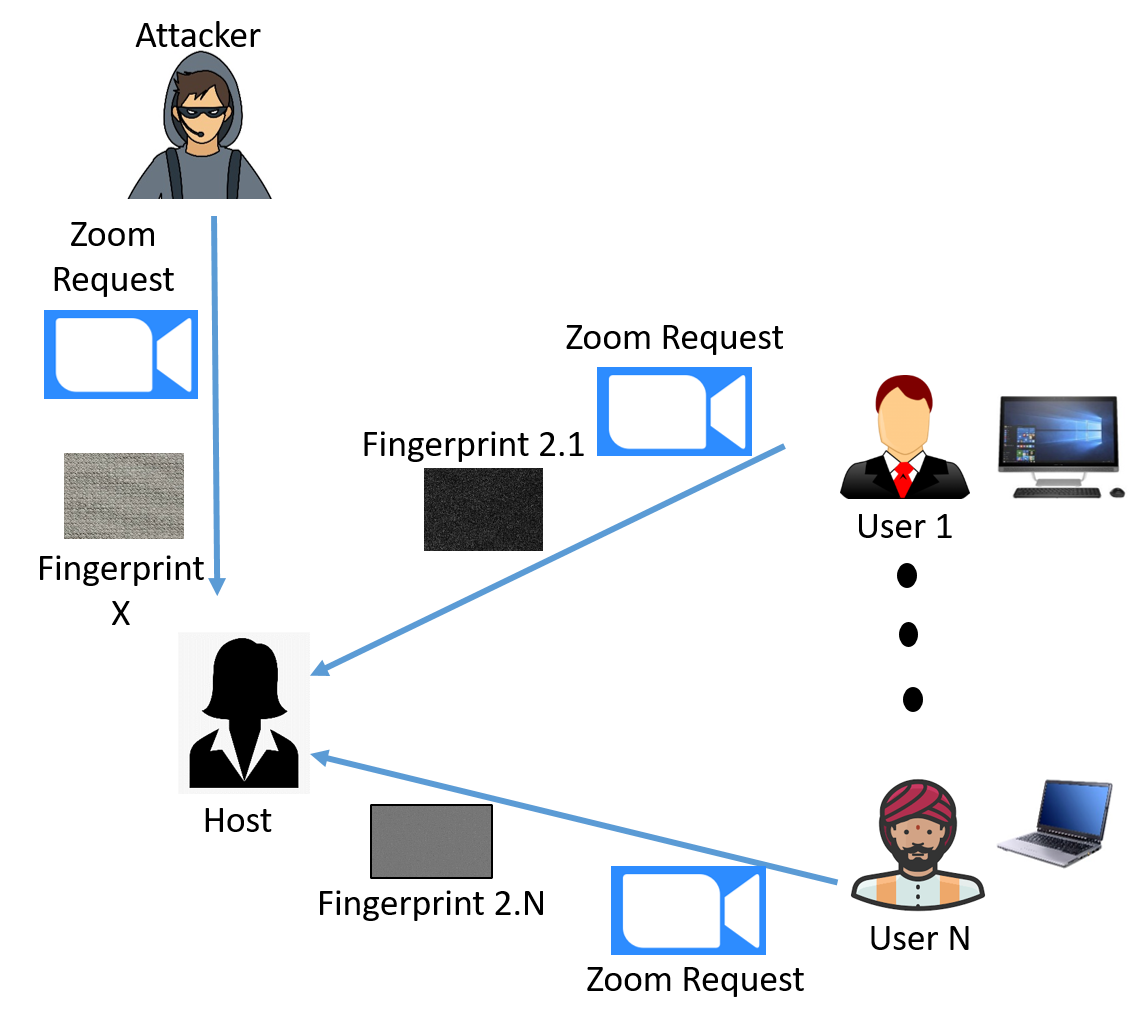}
    \caption{Camera fingerprint matching. 'User 1' to 'User N' are N legitimate participants.}
    \label{fig:Join}
\end{figure}

The proposed authentication method uses both PRNU-based authentication and password-based authentication for making authentication process as seamless as possible. First the PRNU-based authentication is invoked. If a participant is authenticated using this method, she is allowed to Zoom meeting. Otherwise, the password-based authentication is invoked. If the participant can enter a valid password, she is also allowed to join the meeting. Otherwise, the participant is not allowed to the meeting. The PRNU-based authentication is a truly seamless method. In this method, the participant's direct involvement is not required as she is not asked to provide her biometric (e.g., facial expression or fingerprint), or asked to enter a password or respond to a security question. Rather, participant's camera (of the device that the participant uses in Zoom sessions) is authenticated without prompting her for anything. The PRNU-based method does not have $100\%$ true positive rate. Thus a few  legitimate participants will not be authenticated using this method. Theses participants  will only be asked to enter a password. 

Our method has two main steps: Camera fingerprint registration and Camera fingerprint matching. 

Figure~\ref{fig:Regi} provides an overview of the fingerprint registration step. In this step, each legitimate participant of a Zoom meeting must register the fingerprint of her camera with the meeting host. This step is done only once before the start of a meeting. For recurring meetings, registration is also required once before the start of the meeting series. This step can be implemented in a number of ways. A possible way that is more suitable for recurring meetings is to integrate the registration step in a customized Zoom installation app. The host will send the installation app to each participants. During the time of installation, a short video (say, one minute) will be automatically taken (after taking participant's permission) using the camera of the device that is installing the App. Then a camera fingerprint from the video will be computed using the method described in Section~\ref{sec:RelatedeWork1}. The computed fingerprint (e.g., Fingerprint $1.N$ by Participant N in Figure~\ref{fig:Regi}) will be sent to the host. This way of implementing can perfectly suit online teaching setup, where the students can be asked by the educational institute to install a customized Zoom app. Another possible way of implementation is to ask the participants for sending a short video from their camera using a secure communication method (such as secure email). The host can then compute the camera fingerprint of each participant. 

Figure~\ref{fig:Join} provides an overview of the fingerprint matching step. In this step, the authenticity of each participant is checked. This step is performed each time a participant wants to join a Zoom meeting. From initial few seconds of the participant's video, a camera fingerprint is computed. The camera fingerprint (e.g., Fingerprint $2.N$ for Participant N in Figure~\ref{fig:Join}) is then sent to the host. Note that for computing the fingerprint and sending it to the host, no assistance from the participant is required. For each participant, the host matches the registered fingerprint with recently obtained fingerprint (for example, Fingerprint $1.N$ is matched with Fingerprint $2.N$ for Participant N). If the match result is above a threshold, the participant is authenticated, and her Zoom joining request is approved. Otherwise, the user is asked to enter a password that was setup by the meeting host. The current version of Zoom allows the meeting host to setup a password. This password needs to be sent to the participants using other communication methods (such as email or phone).

\section{Experiments}
\label{sec:experiments}

\begin{table}[h]
\centering
\caption{Run time performance of registration and verification of various cameras}
\label{tab:runtime}
\resizebox{1\columnwidth}{!}{
\begin{tabular}{|c|c|c|c|c|c|}
\toprule
\textbf{\begin{tabular}[l]{@{}l@{}}Camera\\ Name \end{tabular}} & \textbf{\begin{tabular}[l]{@{}l@{}}Pixels\\ Size \end{tabular}} & \textbf{\begin{tabular}[l]{@{}l@{}}Register time\\ I-frames (secs) \end{tabular}} & \textbf{\begin{tabular}[l]{@{}l@{}}Register time\\ FP (secs) \end{tabular}} & \textbf{\begin{tabular}[l]{@{}l@{}}Verify time \\ 100 frames \end{tabular}} &
\textbf{\begin{tabular}[l]{@{}l@{}}First frame \\ PCE \end{tabular}} \\
\midrule

Huawei honor 8       & 1280 x 720  & $1.20$ &  $23.84$  &$42.02$ &$178.08$     \\
Samsung S9+          & 1280 x 720  & $3.24$ &  $23.69$  &$40.19$ &$549.42$  \\
Dell Laptop XPS      & 1280 x 720  & $1.71$ &  $23.21$  &$43.21$ &$368.31$   \\
Dell Desktop         & 1280 x 720  & $2.26$ &  $25.68$  &$43.32$ &$2675.7$    \\
HP laptop            & 1280 x 720  & $1.94$ &  $25.00$  &$44.22$ &$2994.3$    \\
Iphone 8 Plus        & 1920 x 1088 & $5.00$ &  $52.86$  &$95.75$ &$155.41$    \\
Dell Laptop XPS      & 1280 x 720  & $2.11$ &  $25.61$  &$43.20$ &$5075.9$    \\
Samsung Note 9       & 1280 x 720  & $2.60$ &  $25.91$  &$40.59$ &$284.99$  \\
Samsung Note 10      & 1280 x 720  & $2.02$ &  $24.05$  &$41.19$ &$5919.3$   \\
Dell Laptop Inspiron & 1280 x 720  & $0.92$ &  $10.57$  &$43.06$ &$5340.2$   \\

\midrule

\end{tabular}
}
\end{table}

The proposed method is experimentally validated by assessing the performance of the PRNU-based method for short videos taken by front cameras of various computing devices (PC, laptop, mobile, tab). In this small scale-experiment, the goal is to study the feasibility of the proposed method.

\begin{figure}
    \centering
    \includegraphics[width=3.3in]{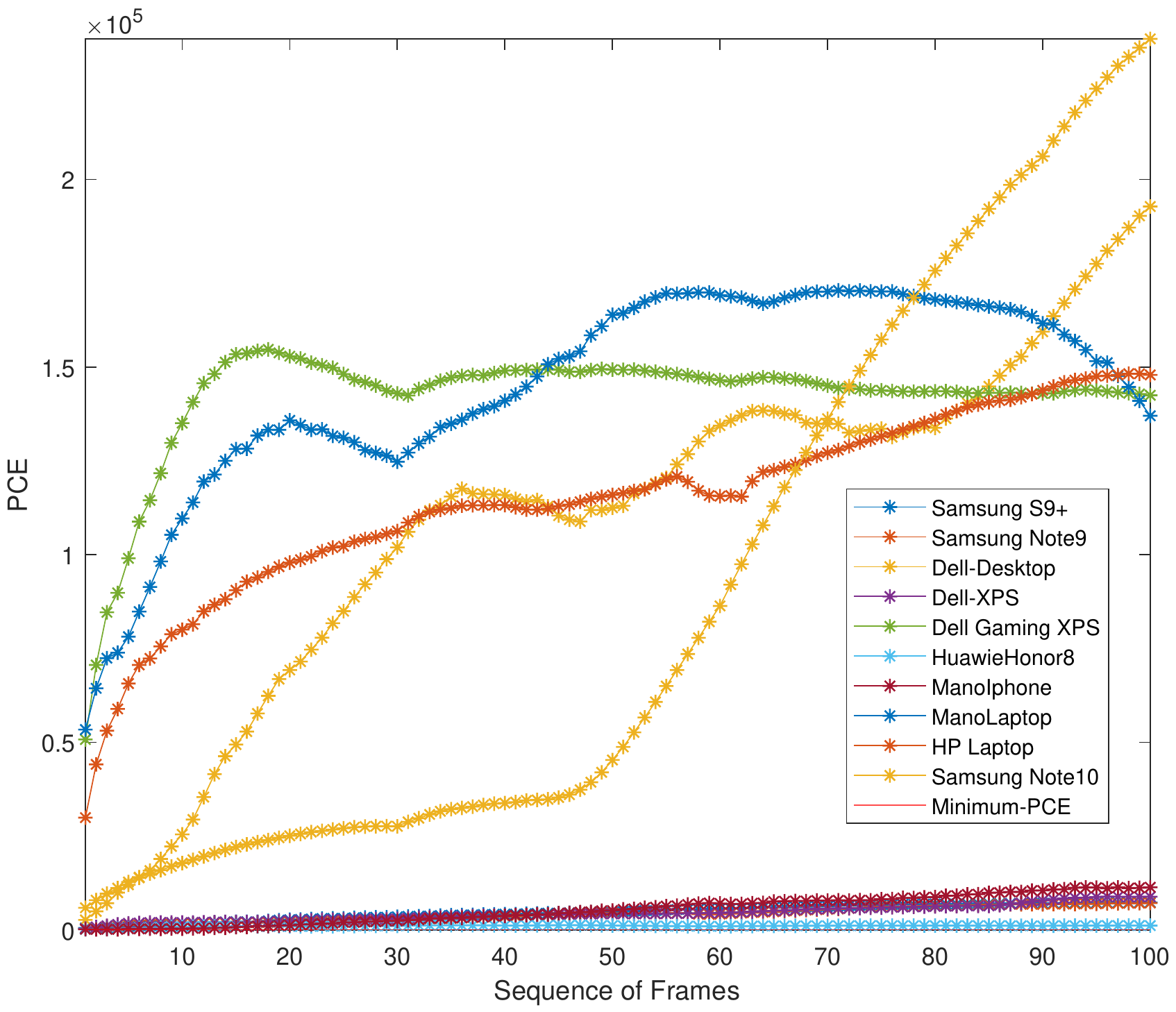}
    \caption{Various Cameras PCE in the first 100 frames of video based on added Noise of subsequent frames}
    \label{fig:my_label14}
\end{figure}

The PRNU-based method (both fingerprint computation and matching) is implemented in MATLAB on a Windows system having $32$ GB RAM, $3.2$ GHz CPU. Videos from $10$ different computing devices were used. Table \ref{tab:runtime} provides the list of computing devices considered for the experiment. From each computing device, a HD video (as Zoom uses HD video \cite{zoom20134tips}) is considered. The videos have not gone through stabilization or out-camera processing (like scaling, cropping, etc.). For computing the camera fingerprint used in the registration step, $60$ i-frames from a $60$ seconds video is used. For computing the camera fingerprint used in the matching step (i.e., query fingerprint), $100$ frames (I, P, and B frames) from a shorter $4$ seconds video is used. This short video is used as for providing a smaller delay (due to PRNU matching) for meeting requests. The PRNU matching is done using PCE with a threshold of $60$.

Figure~\ref{fig:my_label14} shows the performance of the proposed method. Out of $10$ matching, $10$ matching were successful (as the obtained PCE is larger than the threshold $60$) by giving a true positive rate of $100\%$ for this small scale experiment. The achieved true positive rate is by no means generalisable. A large scale experiment is required to find the error rates. However, this small scale experiment shows that the proposed approach works.

Table~\ref{tab:runtime} show the computation cost in computing the query fingerprint and performing matching (\textit{Verify time}) from $100$ frames. This cost is the main contributor to the run-time delay that a user needs to bear due to authentication. This table also reports the computation cost required in the registration phase. This cost, however, is one-time offline cost. Note that, for a smaller video, all the computation costs will be lesser; and for a larger video, these costs will be higher. This trade-off is due to the fact that a lesser number of frames will require lesser number of denoising (which is the main contributor to the delay).

\section{Conclusion and Future Work}
\label{sec:ConclusionAndFutureWork}
Zoom-based online meeting has become very popular due to current coronavirus scenario. Zoom, however, has serious security and privacy issues. One of the main security issue is the poor authentication mechanism, due to which an unauthorized person is able to join a meeting and create disturbances leading to Zoom bombing and Zoom eavesdropping. In this paper, we have proposed our preliminary work towards a seamless authentication scheme. The proposed scheme uses PRNU-based camera authentication method to authenticate a meeting participants. Those who are not authenticated using this method need to provide a password. A small scale experiment shows that the proposed method works as expected. However, a number of improvements need to be done for making the method usable. We have provided some of the possible improvements below. 

\subsection{Future Work}
\textbf{Large Scale Experiment:} A large scale experiment needs to be done for finding better estimates of error rates (e.g., true positive rate and false positive rate). 
\textbf{Protection against attacks:} There can be a number of attacks. For example, if an attacker collects a video of a legitimate participant from another source (e.g., social media), she can impersonate the participant. Techniques need to be developed for withstanding such attacks.
\textbf{Developing an app:} An app implementing the proposed idea needs to be developed. A large scale user study needs to be conducted by asking users (both computer experts and non-computer users) to use the app. This study will provide a better evaluation of the proposed idea in terms of performance, usability, and security.

\section*{Acknowledgement}
This work has been supported by University of Technology MaPS Startup funding 263010-0226628.


\bibliographystyle{IEEEtran}
\bibliography{ZoomAuth}

\end{document}